\documentclass[trackchanges]{aastex63}

\shorttitle{HASSW}
\shortauthors{Huang et al.}

\graphicspath{{./}{figures/}}

\begin{document}

\title{The Temperature Anisotropy and Helium Abundance Features of Alfvénic Slow Solar Wind Observed by \textit{Parker Solar Probe}, \textit{Helios}, and \textit{Wind} Missions}

\correspondingauthor{Jia Huang}
\email{huangjia.sky@gmail.com; huangjia@berkeley.edu}

\author[0000-0002-9954-4707]{Jia Huang}
\affiliation{Space Sciences Laboratory, University of California, Berkeley, CA 94720, USA.}

\author[0000-0001-5030-6030]{Davin E. Larson}
\affiliation{Space Sciences Laboratory, University of California, Berkeley, CA 94720, USA.}

\author[0000-0002-8475-8606]{Tamar Ervin}
\affiliation{Space Sciences Laboratory, University of California, Berkeley, CA 94720, USA.}
\affiliation{Department of Physics, University of California, Berkeley, Berkeley, CA 94720, USA.}

\author[0000-0003-2981-0544]{Mingzhe Liu}
\affiliation{Space Sciences Laboratory, University of California, Berkeley, CA 94720, USA.}

\author{Oscar Ortiz}
\affiliation{Space Sciences Laboratory, University of California, Berkeley, CA 94720, USA.}

\author[0000-0002-7365-0472]{Mihailo M. Martinovi\'c}
\affiliation{Lunar and Planetary Laboratory, University of Arizona, Tucson, AZ 85719, USA.}

\author[0000-0003-1674-0647]{Zhenguang Huang}
\affiliation{Climate and Space Sciences and Engineering, University of Michigan, Ann Arbor, MI 48109, USA.}

\author[0000-0001-8478-5797]{Alexandros Chasapis}
\affiliation{Laboratory for Atmospheric and Space Physics, University of Colorado Boulder, Boulder, CO 80303, USA.}

\author[0000-0003-4109-0770]{Xiangning Chu}
\affiliation{Laboratory for Atmospheric and Space Physics, University of Colorado Boulder, Boulder, CO 80303, USA.}

\author[0000-0001-6673-3432]{B. L. Alterman}
\affiliation{Southwest Research Institute, San Antonio, TX 78238, USA.}

\author[0000-0001-9570-5975]{Zesen Huang}
\affiliation{Department of Earth, Planetary, and Space Sciences, University of California, Los Angeles, CA 90095, USA.}

\author[0000-0001-8495-9179]{Wenwen Wei}
\affiliation{Space Sciences Laboratory, University of California, Berkeley, CA 94720, USA.}

\author[0000-0003-1138-652X]{J. L. Verniero}
\affiliation{Heliophysics Science Division, NASA Goddard Space Flight Center, Greenbelt, MD 20771, USA.}

\author[0000-0002-6849-5527]{Lan K. Jian}
\affiliation{Heliophysics Science Division, NASA Goddard Space Flight Center, Greenbelt, MD 20771, USA.}

\author[0000-0003-3255-9071]{Adam Szabo}
\affiliation{Heliophysics Science Division, NASA Goddard Space Flight Center, Greenbelt, MD 20771, USA.}

\author[0000-0002-4559-2199]{Orlando Romeo}
\affiliation{Space Sciences Laboratory, University of California, Berkeley, CA 94720, USA.}

\author[0000-0003-0519-6498]{Ali Rahmati}
\affiliation{Space Sciences Laboratory, University of California, Berkeley, CA 94720, USA.}

\author[0000-0002-0396-0547]{Roberto Livi}
\affiliation{Space Sciences Laboratory, University of California, Berkeley, CA 94720, USA.}

\author[0000-0002-7287-5098]{Phyllis Whittlesey}
\affiliation{Space Sciences Laboratory, University of California, Berkeley, CA 94720, USA.}

\author[0000-0001-6125-6411]{Samer T. Alnussirat}
\affiliation{Space Sciences Laboratory, University of California, Berkeley, CA 94720, USA.}

\author[0000-0002-7077-930X]{Justin C. Kasper}
\affiliation{BWX Technologies, Inc., Washington DC 20001, USA.}

\author[0000-0002-7728-0085]{Michael Stevens}
\affiliation{Smithsonian Astrophysical Observatory, Cambridge, MA 02138, USA.}

\author[0000-0002-1989-3596]{Stuart D. Bale}
\affiliation{Space Sciences Laboratory, University of California, Berkeley, CA 94720, USA.}
\affiliation{Department of Physics, University of California, Berkeley, Berkeley, CA 94720, USA.}

\begin{abstract}
Slow solar wind is typically characterized as having low Alfvénicity, but the occasional occurrence of highly Alfvénic slow solar wind (HASSW) raises questions about its source regions and evolution. In this work, we conduct a statistical analysis of temperature anisotropy and helium abundance in HASSW using data from PSP within 0.25 AU, Helios between 0.3 AU and 1 AU, and Wind near 1 AU. Our findings reveal that HASSW is prevalent close to the Sun, with PSP observations displaying a distinct ``U-shaped" Alfvénicity distribution with respect to increasing solar wind speed, unlike the monotonic increase trend seen in Helios and Wind data. This highlights a previously unreported population of unusually low speed HASSW, which is found in both sub-Alfvénic and super-Alfvénic regimes. The observed decreasing overlap in temperature anisotropy between HASSW and fast solar wind (FSW) with increasing heliocentric distance suggests different underlying heating processes. Additionally, HASSW exhibits two distinct helium abundance populations, particularly evident in PSP data, with generally higher helium abundance compared to less Alfvénic slow solar wind. Moreover, the decreasing overlap in temperature anisotropy versus helium abundance distributions between HASSW and FSW with decreasing radial distance implies that not all HASSW originates from the same source region as FSW. 
\end{abstract}

\keywords{Alfvénic slow solar wind, temperature anisotropy, helium abundance, radial evolution, origin}

\section{Introduction} \label{sec:intro}
Solar wind is an ionized plasma that flows out from the Sun, consisting of protons, electrons, alpha particles, and some heavy ions \citep[e.g.][]{priest-2014}. It can be categorized based on various parameters or characteristics, such as speed, inferred source region, and Alfvénicity. Among these, solar wind speed is a simple yet effective criterion, classifying it into slow solar wind (SSW; $V_p < 450$ km/s) and fast solar wind (FSW; $V_p > 450$ km/s), which generally exhibit distinct plasma characteristics and originate from different source regions \citep[e.g.][]{wang-2000, mccomas-2013}. 

The compositional measurements of helium particles and heavy ions provide critical insights into identifying the source regions of different solar winds \citep{geiss-1995, bochsler-2007}. The freezing-in temperature, inferred from the charge states of heavy ions, reflects the coronal electron temperature. Because the ionization and recombination processes of heavy ions are generally balanced at about 1.2-3.5 solar radii ($R_S$), and this temperature frozen in as the solar wind propagates outward into a collisionless regime \citep{burgi-1986, zhao-2009, huang-2018}. Elemental abundance ratios are associated with the first ionization potential (FIP) effect, which results from processes occurring near the large temperature gradients at the base of the solar transition region \citep{geiss-1982, raymond-1999}. These compositional characteristics are preserved as the solar wind propagates beyond a certain height from the Sun, thereby enabling a confident linkage between in situ solar wind properties and their source regions \citep{kasper-2007}. SSW typically shows higher charge states and elemental abundance ratios of heavy ions compared to FSW, indicating that SSW likely originates from regions with higher electron temperature and larger long-lived magnetic loops \citep{fisk-1999, fisk-2001}, i.e. closed magnetic field regions such as helmet streamers \citep{suess-2009, huang-2023SBSW}, pseudostreamers \citep{crooker-2014, huang-2016a}, active regions \citep{kasper-2007, brooks-2015}, small coronal holes and coronal hole boundaries \citep{higginson-2017, wang-2017, bale-2019}. In contrast, FSW comes from open magnetic field regions, primarily coronal holes and their associated regions \citep{tu-2005, cranmer-2009, mccomas-2013}. In addition, the helium-to-proton abundance ratio ($A_{He} = N_\alpha/N_p \times 100\%$, where $N_\alpha$ and $N_p$ are the number densities of helium particles and protons, respectively) is typically depleted in the vicinity of heliospheric current sheet (HCS) but enhanced in FSW and magnetic clouds in comparison with ambient solar wind, implying that the depletion may originate from closed field regions of helmet streamer \citep{gosling-1981, suess-2009}. Moreover, the variation of $A_{He}$ with solar activity in SSW indicates that the depleted $A_{He}$ (helium-poor population, $A_{He}\approx1.5\%$) during solar minimum likely originate from helmet streamer, whereas the enhanced $A_{He}$ (helium-rich population, $A_{He}\approx4.5\%$) during solar maximum may come from active regions \citep{kasper-2007, kasper-2012, alterman-2018}.

The origins of SSW remain one of the major unsolved problems in heliospheric physics \citep{antiochos-2011}. Addressing this problem is challenging due to the unexpectedly broad latitudinal distribution of SSW, particularly during solar maximum \citep{wang-2000}, and its significantly more dynamic temporal and spatial variations compared to FSW \citep{abbo-2016}. Alfvénicity, which measures the correlation between fluctuations in the components of the magnetic field and solar wind velocity, exemplifies these variable characteristics. Typically, FSW exhibits high Alfvénicity, while SSW shows low Alfvénicity. However, high Alfvénicity SSW (HASSW) was first identified in Helios observations at around 0.3 AU \citep{marsch-1981}, suggesting that this type of SSW shares similar features with FSW, despite its slower speed. Multi-event studies of HASSW \citep[e.g., ][]{dAmicis-2015, dAmicis-2018, perrone-2020} further indicate that HASSW shows nearly identical charge states ratios to FSW, and its proton temperature anisotropy ($T_\perp/T_\parallel$) also displays an anisotropic feature similar to FSW ($T_\perp/T_\parallel > 1$) rather than the isotropic state in regular SSW ($T_\perp/T_\parallel \sim 1$), implying that HASSW and FSW share similar features on both macro- and micro-scales. Generally, temperature anisotropy is defined as the difference in temperature between the direction perpendicular to ($T_\perp$) and the direction parallel to ($T_\parallel$) the background magnetic field. Therefore, HASSW and FSW are likely originated from the same source region, i.e. coronal holes. An investigation of radial variations in temperature anisotropy and Alfvénicity using Helios data \citep{stansby-2019} reveals distinctly separated anisotropic and isotropic temperature profiles around 0.3 AU, with the anisotropic Alfvénic wind arising from the central regions of coronal holes, while the isotropic Alfvénic wind coming from active regions or coronal hole boundaries, and the isotropic non-Alfvénic wind from small scale transients.

In the solar wind, Alfvénicity decreases with distance \citep{roberts-1987, bruno-2007} due to multiple factors, including solar wind interactions, velocity shears, and turbulence evolution \citep{bruno-2006, dAmicis-2018, stansby-2019}. Therefore, investigating the nature of HASSW close to the Sun, prior to any significant decay in Alfvénicity, is crucial. Additionally, a statistical study can provide a more comprehensive understanding of HASSW than multi-event analyses. The Parker Solar Probe (PSP) mission \citep{fox-2016}, designed to make unprecedentedly close approaches to the Sun, offers a unique opportunity for such investigations. Initial PSP observations revealed that HASSW emerges from a small equatorial coronal hole at 36 to 54 $R_S$ \citep{bale-2019}. More recently, \citet{ervin-2024comp} traced an interval of HASSW back to the overexpanded boundary of a coronal hole, and \citet{ervin-2024sasw} further found that HASSW likely originates from both coronal-hole-like structures and non-coronal-hole structures based on a multi-event study of the sources and characteristics of the HASSW. With these unique PSP observations, we statistically compare the properties of HASSW with those observed by Helios and Wind at greater heliocentric distances. This approach allows us to better identify the differences between HASSW and FSW, and thereby determine the source regions of HASSW from a statistical perspective.

\section{Data} \label{sec:data}

\begin{figure}
\epsscale{0.8}
\plotone{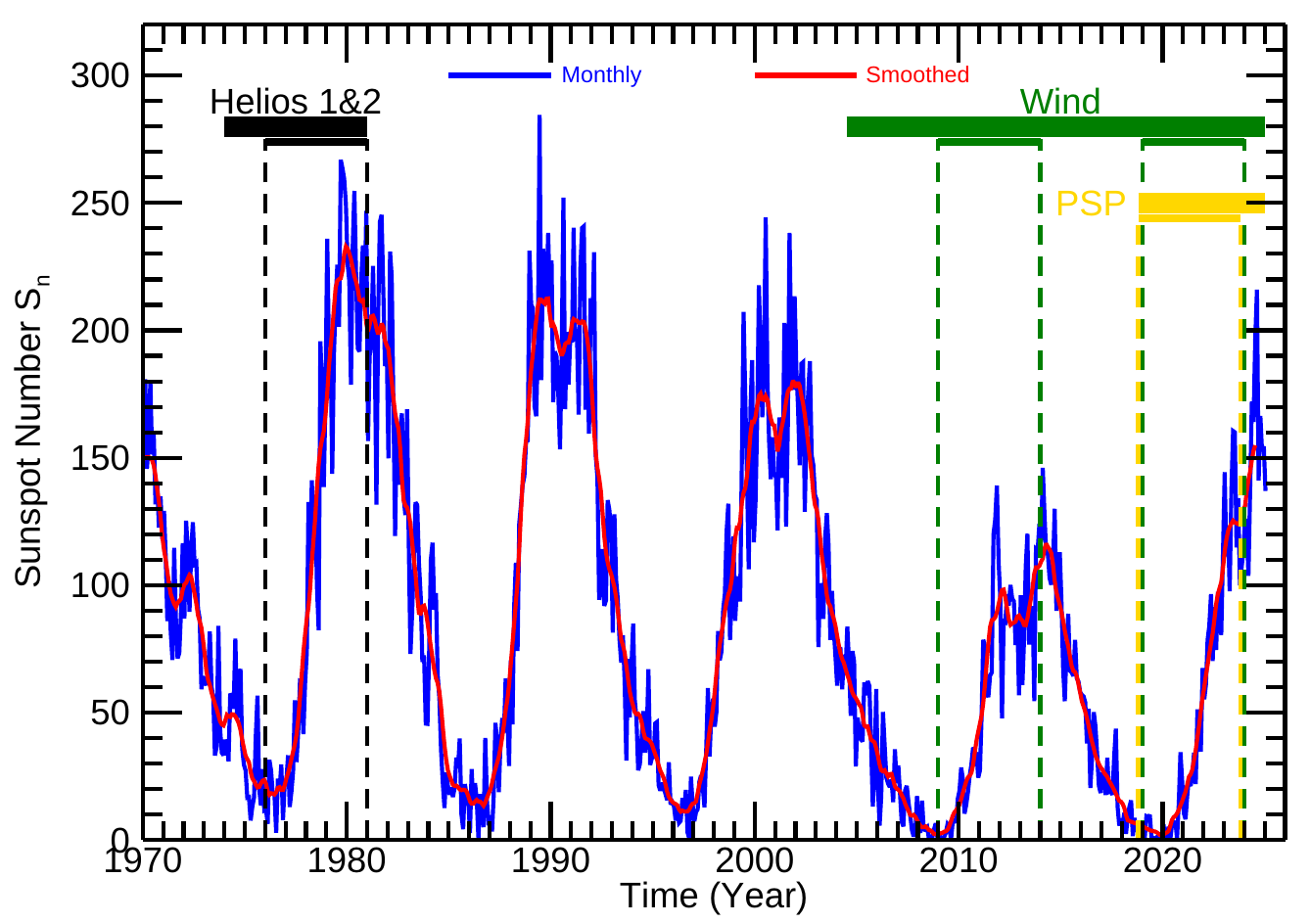}
\caption{Sunspot number and spacecraft data coverage. The monthly mean (blue line) and 13-month smoothed monthly (red line) total sunspot number data are from \citet{sunspots}. The color bands mark the time coverage for the shown missions, and the vertical dashed lines indicate the time range of selected data in this study. } \label{fig:sunspot}
\end{figure}

In this work, we utilize data from PSP below 0.25 AU, Helios between approximately 0.3 AU and 1 AU, and Wind around 1 AU, covering a heliocentric distance range from 0.053 AU to about 1 AU.

For the PSP data used in this work, the Solar Wind Electrons, Alphas, and Protons (SWEAP) instrument suite \citep{kasper-2016} provides the proton and alpha measurements, while the FIELDS instrument suite \citep{bale-2016} offers the magnetic field data. The plasma data have a time resolution of up to 0.437 s, whereas the magnetic field data are down-sampled from a much higher time cadence to match the plasma data. Proton temperature components for the first two encounters are derived using the measurements from FIELDS and Solar Probe Cup of SWEAP \citep[SPC;][]{case-2020} with the method described in \citet{huang-2020}, whereas they are either directly retrieved from the measurements of Solar Probe ANalyzer for Ions of SWEAP \citep[SPAN-I;][]{livi-2022} or obtained through the bi-Maxwellian fitting to the proton spectra observed by SPAN-I since encounter 4. Encounter 3 is excluded from this study due to data gap during perihelion. The plasma dataset is further cleaned based on several criteria: 
\begin{itemize}
    \item[(1)] SPC measurements with poor data quality flags are excluded for the first two encounters. 
    \item[(2)] Since encounter 4, the primary velocity distribution functions of both proton and alpha particles should fall within the field of view of SPAN-I . Due to obstructions by the heat shield in front of the spacecraft, we specifically require that the solar wind flow angle remains below an azimuthal angle of approximately 165° in the instrument coordinate, with the measured energy fluxes of both proton and alpha particles peaking in at least the second-to-last azimuthal angle. 
    \item[(3)] SPAN-I fits data are discarded when the $\chi^2$ per degree of freedom, which measures the goodness of fits, is excessively large. 
    \item[(4)] The deviation of the measured proton and alpha densities from the quasi-thermal noise \citep[QTN;][]{moncuquet-2020, liu-2021} electron density ($N_e$) remains within a reasonable range under the assumption of a neutral plasma state, i.e. $|N_p+2N_\alpha-N_e|/N_e < 50\%$. 
    \item[(5)] The alpha data are further required to be within reasonable and relatively flexible limits, i.e. $0.1\% < A_{He} < 20\%$ and $|V_\alpha-V_p|/V_A < 3.0$, where $V_\alpha$, $V_p$, and $V_A$ are the alpha speed, proton speed, and local Alfvén speed, respectively. 
\end{itemize}

The Helios 1\&2 dataset is from a recent reprocessing of the velocity distribution function (VDF) measurements, and it provides fits for a proton core, proton beam, and helium component for nearly the entire Helios 1 \& 2 missions by assuming bi-Maxwellian VDFs for all components \citep{durovcova-2019Helios}.  

For the Wind data used in this work, the SWE Faraday cup measures the reduced VDFs of solar wind proton and helium along 40 angles every 92 s \citep{ogilvie-1995}, and the MFI provides the magnetic field data \citep{lepping-1995, koval-2013}. The proton temperature anisotropies are derived by fitting the spectra with a convected bi-Maxwellian function \citep{kasper-2007}. Additionally, we note that Wind has resided at the Lagrange 1 point since June 2004. 

As shown in Figure \ref{fig:sunspot}, the time coverage of the three missions is highlighted by color bands. Specifically, Helios 1 \& 2 data from 1974 to 1981 predominantly cover the ascending phase of solar cycle 21, Wind data at Lagrange 1 point from June 2004 encompass the entire solar cycle 24, as well as the descending phase of solar cycle 23 and ascending phase of solar cycle 25, whereas PSP data since 2018 primarily cover the ascending phase of solar cycle 25. This figure also indicates that solar activity during the Helios observations was significantly higher than during the periods when Wind and PSP were in operation. To remove the potential bias in the results caused by solar cycle dependency, we select measurements from the ascending phase of the solar cycle for comparison. The chosen time intervals are marked by vertical dashed lines in the figure. Specifically, Helios data from 1976 to 1981, and Wind data from 2009 to 2014 and from 2019 to 2024, are selected. For PSP data, we use measurements from encounters 1 to 17 and apply the above listed criteria to clean the dataset, focusing on approximately 13 days around the perihelion in each encounter. 

\section{Observations} \label{sec:obser}

\subsection{Overview  \label{sec:overview}}

\begin{figure}
\epsscale{1.}
\plotone{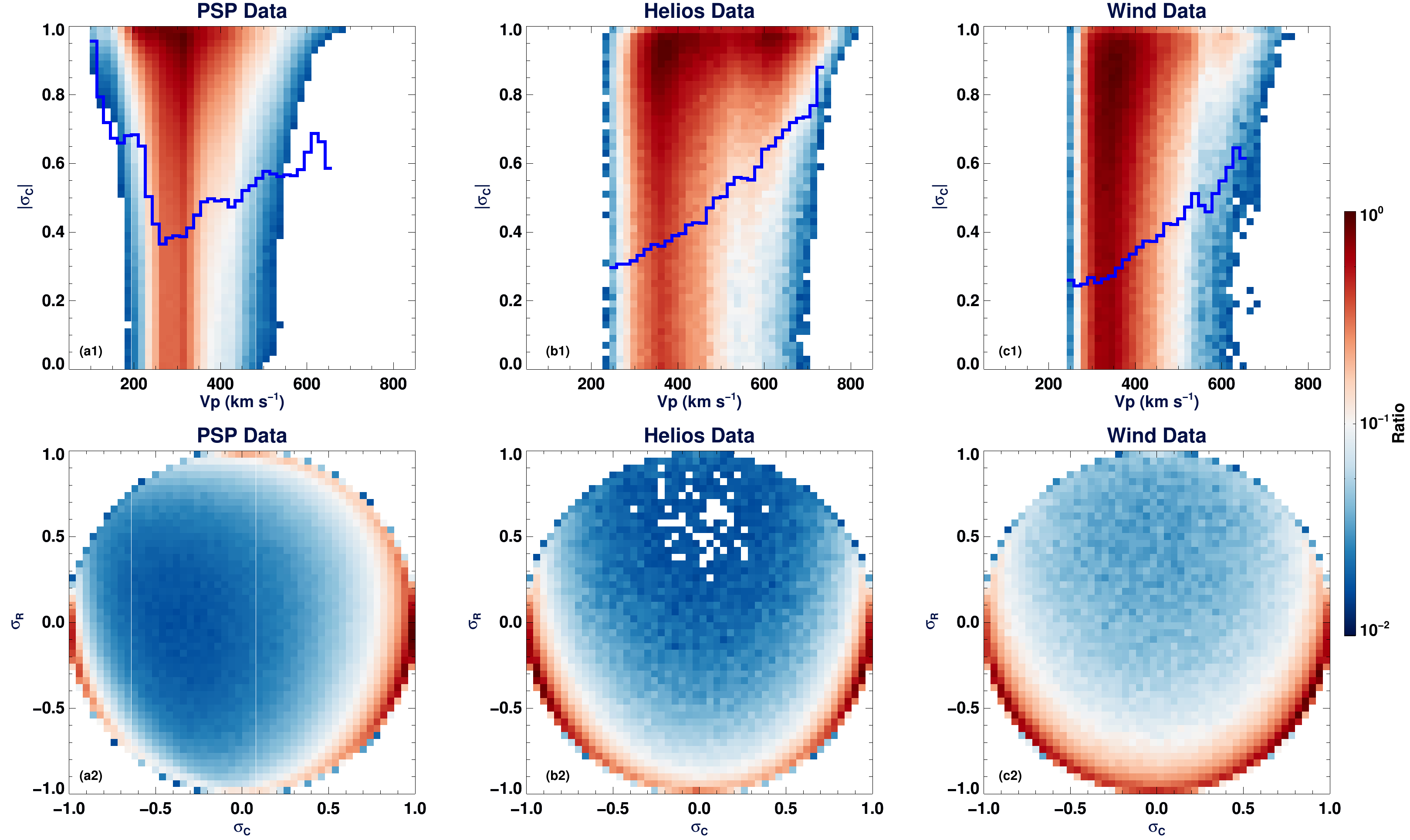}
\caption{Absolute cross helicity variations at different heliocentric distances. The upper panels show the absolute cross helicity $|\sigma_C|$ variations with solar wind speed $V_p$. The blue histogram lines indicate the ratio of high Alfvénicity ($|\sigma_C| > 0.75$) solar wind at each speed bin. The lower panels show the cross helicity $\sigma_C$ variations with residual energy $\sigma_R$. The color in each grid indicates the occurrence ratio of measurements that is normalized by the maximum number of measurements in all grids. } 
\label{fig:sgC}
\end{figure}

The Alfvénicity is evaluated by the normalized cross helicity ($\sigma_C$) and normalized residual energy ($\sigma_R$), which are derived from the trace power spectra of the velocity fluctuation $\delta\textbf{v}$, the magnetic field fluctuation $\delta\textbf{b}$, and Els\"asser variables $\textbf{z}^\pm$. These trace power spectra are denoted as $E^v$, $E^b$, and $E^\pm$, respectively \citep{wicks-2013, chen-2013}. The Els\"asser variables are defined as $\textbf{z}^\pm = \delta\textbf{v} \pm \delta\textbf{b}/\sqrt{\mu_0\rho_0}$, where $\mu_0$ is the vacuum magnetic permeability and $\rho_0$ is proton mass density. The residual energy is calculated to measure the difference between kinetic energy ($E^v$) and magnetic energy ($E^b$). They are defined as: 

\begin{equation}
\label{eq:sgc}
\sigma_C = \frac{E^+ - E^-}{E^+ + E^-} = \frac{2<\delta\textbf{v}\cdot\delta\textbf{b}>}{<\delta\textbf{v}^2>+<\delta\textbf{b}^2>}
\end{equation}

\begin{equation}
\label{eq:sgr}
\sigma_R = \frac{E^v - E^b}{E^v + E^b} = \frac{<\delta\textbf{v}^2>-<\delta\textbf{b}^2>}{<\delta\textbf{v}^2>+<\delta\textbf{b}^2>}
\end{equation}

In general, $\textbf{z}^+$ and $\textbf{z}^-$ represent forward and backward propagating modes with respect to the mean magnetic field orientation. Alfvénic fluctuations are characterized by high normalized cross helicity ($|\sigma_C|\sim1$) and low normalized residual energy ($|\sigma_R|\sim0$). A high $|\sigma_C|$ indicates that energy is predominantly in either $\textbf{z}^+$ or $\textbf{z}^-$ mode, whereas a low $|\sigma_R|$ suggests equipartitioned kinetic and magnetic fluctuation energies. These are features of Alfvén waves, and the $\sigma_C$ values of +1 and -1 correspond to pure Alfvén waves propagating anti-sunward and sunward, respectively. 
Previous studies have used various time scales, ranging from several minutes to some hours, to calculate cross helicity \citep[e.g.][]{wicks-2013, jagarlamudi-2019}. In this work, we calculate the cross helicity every six hours, which represents approximately several correlation times as measured by PSP \citep{chen-2020, parashar-2020, cuesta-2022tur}, Helios \citep{cuesta-2022, cuesta-2022tur}, and Wind \citep{jagarlamudi-2019, cuesta-2022tur}. In addition, PSP has observed prevalent switchbacks, which are highly Alfvénic structures, in the inner heliosphere \citep{bale-2019, kasper-2019, huang-2023SWBOrigin, huang-2023SWBTemp}. However, \citet{mcmanus-2020} suggests that $\sigma_C$ is unaffected by switchbacks, thus we will not exclude them from our study. But the strong interplanetary coronal mass ejection beginning on 2022 September 5 \citep[e.g.,][]{romeo-2023} was excluded. 

Figure \ref{fig:sgC} presents an overview of the absolute cross helicity variations as a function of solar wind speed (upper panels) and residual energy (lower panels) across different heliocentric distances. From left to right, the panels show the data from PSP below 0.25 AU, Helios between 0.3 and 1.0 AU, and Wind around 1 AU. The color in each grid represents the occurrence ratio, calculated as the number of measurements in that grid normalized by the maximum number of measurements across all grids, while the overlapping blue histograms in the upper panels indicate the fraction of high Alfvénicity solar wind ($|\sigma_C| > 0.75$) within each speed bin. 
The top panels indicate that high Alfvénicity populations exist across all solar wind speeds, though their proportions vary. Panels (b1-c1) show that the fraction of high Alfvénicity population monotonically decreases with decreasing solar wind speed, dropping to about 260 km/s as indicated by the blue histograms. Specifically, 46.4\% of Helios data are classified as high Alfvénicity solar wind, with 16.3\% identified as HASSW. For Wind observations, these numbers are 33.6\% and 19.7\%, respectively. Panels (b2-c2) further suggest that high Alfvénicity populations generally exhibit nearly balanced kinetic and magnetic energy ($\sigma_R \sim 0$), implying predominantly pure Alfvénic fluctuations close to the Sun; however, non-Alfvénic populations show a dominance of magnetic energy ($\sigma_R \sim -1$). 

Panel (a1) displays PSP observations of Alfvénicity as a function of solar wind speed, revealing a ``U-shaped” trend as indicated by the blue histogram line. For relatively high speed solar wind, the high Alfvénicity population decreases with decreasing solar wind speed, reaching a similar minimum around 260 km/s as seen in Helios and Wind data. However, for the solar wind speeds slower than 260 km/s, the high Alfvénicity population increases with decreasing speed until around 100 km/s, which might be the slowest solar wind recorded and it dominates about 0.12\% of total measurements. The origin of highly Alfvénic very slow solar wind ($<$ 260 km/s) is currently under investigation. Additionally, the PSP data show that 44.1\% of the solar wind is a high Alfvénicity population, with 36.1\% classified as HASSW, meaning that HASSW constitutes 81.8\% of the high Alfvénicity population, a proportion significantly higher than those observed by Helios (35.1\%) and Wind (58.6\%). This result indicates that HASSW is considerably more prevalent near the Sun compared to farther distances and dominates the high Alfvénicity population in the inner heliosphere. Furthermore, Helios observations indicate the largest high Alfvénicity population, but the smallest HASSW percentage compared to PSP and Wind data, implying potential radial and solar cycle dependencies in the high Alfvénicity population.

\subsection{Temperature Anisotropy \label{sec:temp}}

\begin{figure}
\epsscale{1.}
\plotone{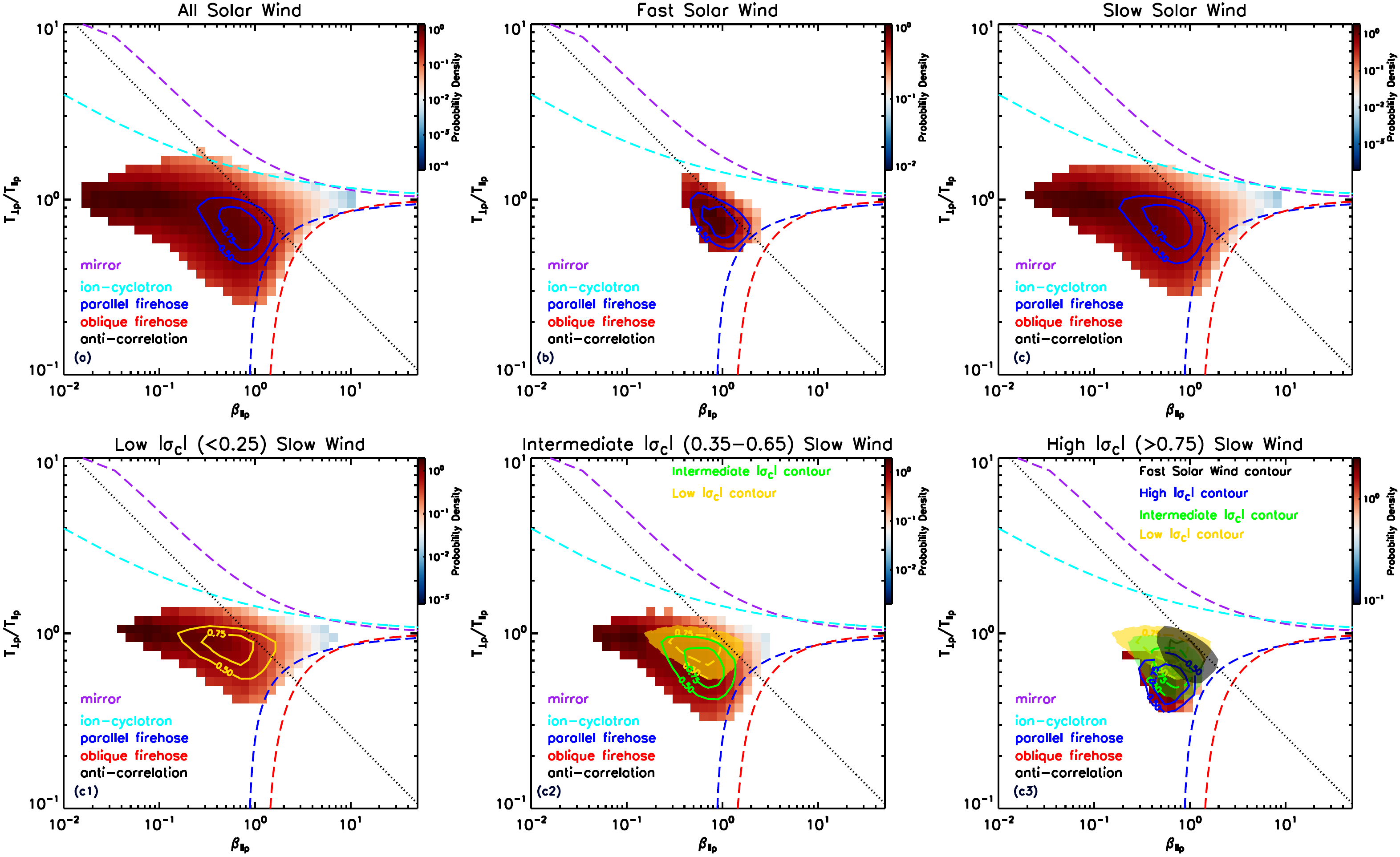}
\caption{Wind observations of the probability density variations of temperature anisotropy versus parallel plasma beta in different solar wind streams. Panels (a) to (c) show the variations in all solar wind, fast solar wind, and slow solar wind, respectively. Panels (c1) to (c3) present the variations in low Alfvénicity slow wind ($|\sigma_C|<0.25$), intermediate Alfvénicity slow wind ($0.35\leq|\sigma_C|\leq0.65$), and high Alfvénicity slow wind ($|\sigma_C|>0.75$). The purple, cyan, blue, and red dashed lines in each panel indicate the mirror, ion-cyclotron, parallel firehose, and oblique firehose instabilities, respectively. The black dotted line represents the anti-correlation between temperature anisotropy and parallel plasma beta given by Eq. \ref{eq:anti}. In each panel, the contours indicate 50\% and 75\% measurement levels. In panels (c2) and (c3), the shaded contours of other types of solar wind are overlapped for comparison, as indicated by the legend. } \label{fig:TaniSSW_wind}
\end{figure}

\begin{figure}
\epsscale{1.}
\plotone{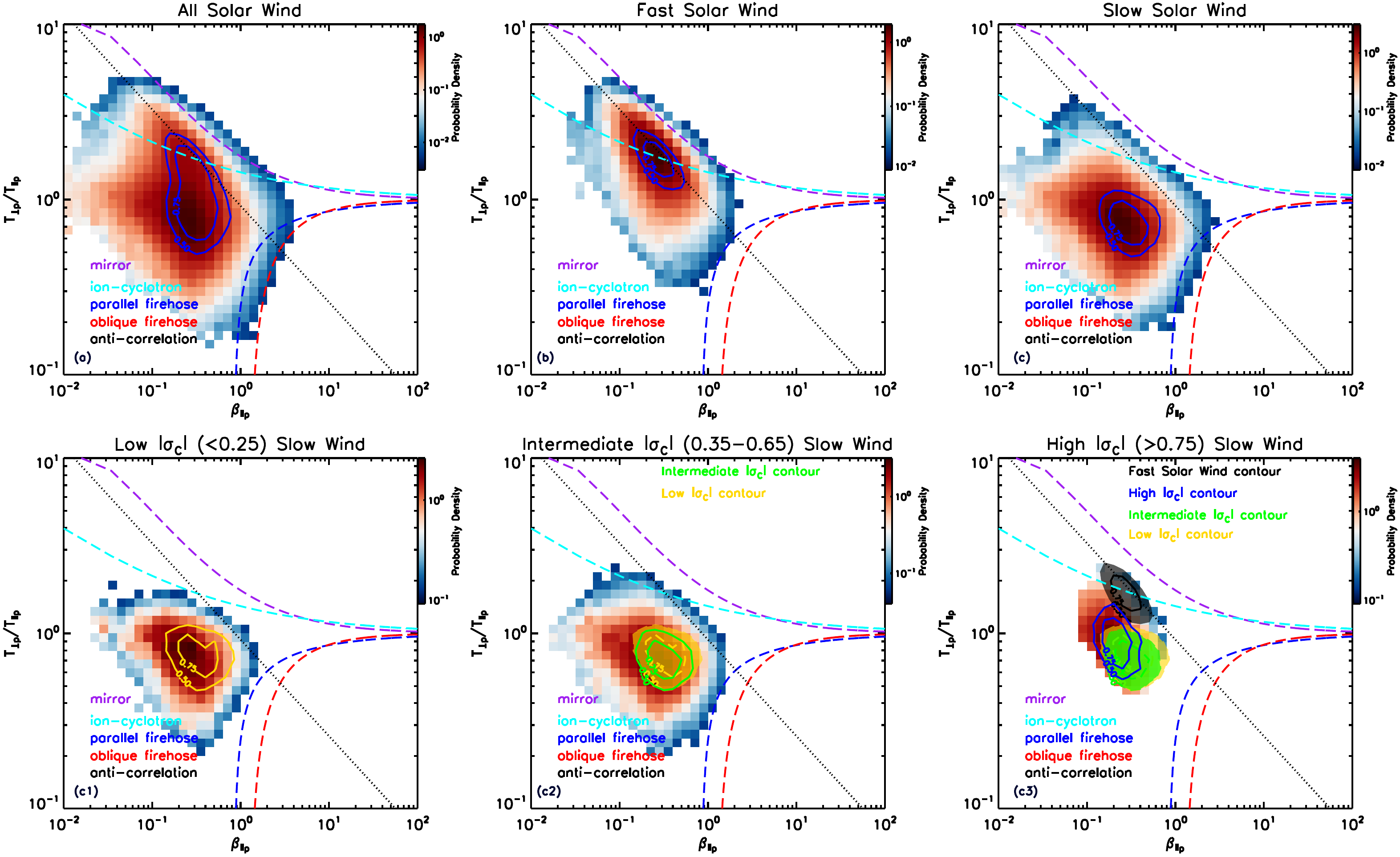}
\caption{Helios observations of the probability density variations of temperature anisotropy versus parallel plasma beta in different solar wind streams. The format is the same as Figure \ref{fig:TaniSSW_wind}. } \label{fig:TaniSSW_Helios}
\end{figure}

\begin{figure}
\epsscale{1.}
\plotone{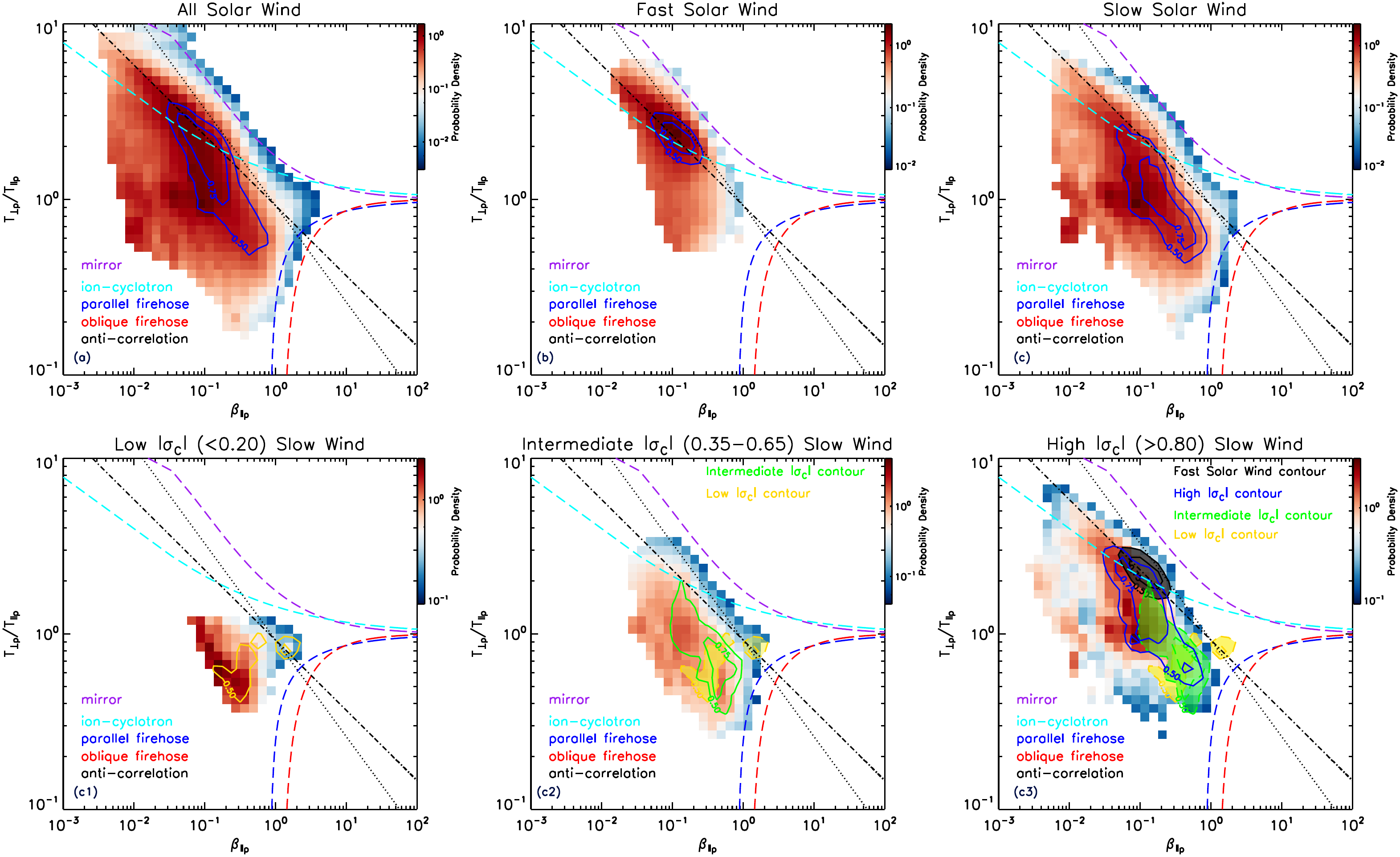}
\caption{PSP observations of the probability density variations of temperature anisotropy versus parallel plasma beta in different solar wind streams. The format is the same as Figure \ref{fig:TaniSSW_wind}, except for the dot-dashed line representing a newly fitted anti-correlation. We further use a slightly stricter criterion to select low Alfvénicity (i.e., $|\sigma_C|<0.20$) and high Alfvénicity (i.e., $|\sigma_C|>0.80$) slow solar wind. } \label{fig:TaniSSW_PSP}
\end{figure}

In this section, we compare the temperature anisotropy features of different types of solar wind using data from Wind, Helios, and PSP, to determine whether HASSW exhibits features more similar to FSW or low/intermediate Alfvénicity SSW.

Figure \ref{fig:TaniSSW_wind} presents Wind observations of temperature anisotropy as a function of parallel plasma beta, $\beta_{\parallel p} = 2\mu_0 N_p k_B T_{\parallel p}/B^2$, where $B$ is magnetic field strength and $k_B$ is the Boltzmann constant. The color represents the probability density in each bin, calculated using the method described by \citet{maruca-2011}: $p=n/(N\ \Delta\beta_{\parallel p}\ \Delta(T_{\perp p}/T_{\parallel p}))$, where n is the number of data points in the bin, N is the total number of data points, and $\Delta\beta_{\parallel p}$ and $\Delta(T_{\perp p}/T_{\parallel p})$ are the bin widths along each axis. The purple, cyan, blue and red dashed lines in each panel indicate the mirror, ion-cyclotron, parallel firehose, and oblique firehose instabilities, respectively, with the thresholds from \citet{hellinger-2006}. The black dotted line represents the observed anti-correlation between $T_{\perp p}/T_{\parallel p}$ and $\beta_{\parallel p}$, given by: 
\begin{equation}
\label{eq:anti}
    \frac{T_{\perp p}}{T_{\parallel p}} = \frac{a}{\beta_{\parallel p}^{b}}
\end{equation}
where a$\simeq$1.16 and b$\simeq$0.55. This anti-correlation was first derived for the core part of the proton distribution in the fast wind with Helios observations, and it is formed by the resonant interaction between ion-cyclotron waves and protons as described by the quasi-linear theory of pitch angle diffusion \citep{marsch-2004}. As the heliocentric distance increases, the anti-correlation breaks down (around 0.9 AU) due to the solar wind expanding along a marginal stability path that corresponds to the equilibrium between the adiabatic expansion and the firehose saturation; however, SSW does not exhibit this relationship probably due to the slow wind irregularities \citep{matteini-2007}. 

In Figure \ref{fig:TaniSSW_wind}, panels (a) to (c) present the temperature anisotropy distributions for all solar wind, FSW ($V_p > 500$ km/s), and SSW ($V_p < 400$ km/s), respectively, with the blue contours indicating the 50\% and 75\% measurement levels. Here, the solar wind speed thresholds are adjusted to exclude the likely mixed slow and fast solar wind between 400 km/s and 500 km/s. 
FSW in panel (b) exhibits a distribution with an approximately downward-convex quasi-oval shape, which generally aligns with (but, slightly deviates from) the anti-correlation model. Conversely, SSW in panel (c) shows a roughly upward-convex quasi-oval shaped distribution that, as expected, does not follow the anti-correlation. Note that \textit{a} in Equation \ref{eq:anti} is adjusted to 0.92 to better match with the Helios observations shown in Figure \ref{fig:TaniSSW_Helios}. 
Furthermore, the distributions of low ($|\sigma_C|<0.25$), intermediate ($0.35\leq|\sigma_C|\leq0.65$), and high ($|\sigma_C|>0.75$) Alfvénicity SSW are shown in panels (c1-c3), with the contours of other types of solar wind overplotted in panels (c2-c3) for comparison. It is evident that low $|\sigma_C|$ SSW is dominated by isotropic temperatures, while higher $|\sigma_C|$ SSWs exhibit more anisotropic temperatures with $T_{\perp p}/T_{\parallel p} < 1$. Panel (c3) further illustrates that the main distribution shifts downward in the $T_{\perp p}/T_{\parallel p}$ -- $\beta_{\parallel p}$ space from low (yellow contour) to high (blue contour) $|\sigma_C|$ SSW, moving farther away from that of FSW (black shaded contour). This result suggests that HASSW probably undergoes different heating processes compared to FSW, implying that some of the HASSW may not originate from the same source region as FSW. This contradicts with previous results from multi-event studies, which found that HASSW exhibits similar microphysical states to FSW, but deviates from those of regular SSW \citep[e.g.][]{dAmicis-2015, dAmicis-2018}.

Figure \ref{fig:TaniSSW_Helios} displays the temperature anisotropy distributions using Helios data, following the same format as Figure \ref{fig:TaniSSW_wind}. Panels (a-c) indicate two distinct populations in the temperature anisotropy distributions: one corresponding to FSW, which is anisotropic ($T_{\perp p}/T_{\parallel p} > 1$) and aligns well with the anti-correlation model as shown in panel (b), and the other to SSW, which is nearly isotropic as indicated in panel (c). Panels (c1-c3) further compare the distributions in low, intermediate, and high $|\sigma_C|$ SSWs, indicating a similar signature across the different Alfvénicity SSWs. However, the distribution of HASSW (blue contour) is closer to that of FSW (black shaded contour). This result suggests, once again, that the HASSW observed by Helios does not significantly deviate from the regular SSW with low and intermediate Alfvénicities. 
 
Figure \ref{fig:TaniSSW_PSP} shows the temperature anisotropy distributions using PSP observations, following the same format as Figure \ref{fig:TaniSSW_wind}. In this analysis, we slightly adjust the solar wind speed threshold to distinguish FSW ($V_p > 450$ km/s) and SSW ($V_p < 350$ km/s), taking solar wind acceleration into account \citep{shi-2022acceleration, halekas-2023}. The thresholds are approximately determined from the results shown in Figure \ref{fig:sgC}, assuming that PSP observes a similar percentage of high Alfvénicity populations as Helios at these specific speeds, which is similar to the method presented in \citet{ervin-2024sasw} and the authors instead use Wind data for comparison. Additionally, we apply slightly stricter thresholds to select low ($|\sigma_C| < 0.20$) and high ($|\sigma_C| > 0.80$) Alfvénicity SSWs, because PSP observes prevalent Alfvénic solar wind in the inner heliosphere. 
Panels (a-c) also indicate two populations, with one corresponding to FSW and the other to SSW. However, while the FSW distribution still exhibits an anti-correlation, it does not align with the anti-correlation derived from Helios data as indicated by the black dotted line. Instead, it follows a newly fitted anti-correlation, as indicated by the black dot-dashed line, with the free parameters a$\simeq$0.92 and b$\simeq$0.40 in Equation \ref{eq:anti}. This suggests that temperature differences between the perpendicular and parallel directions become more balanced closer to the Sun, potentially due to the stronger perpendicular heating and weaker parallel cooling effects in the near-Sun environment as indicated by \citet{huang-2020}. Moreover, panels (c1-c3) reveal that the main distribution shifts upward from low (yellow contour) to high (blue contour) $|\sigma_C|$ SSW and moves closer to that of the FSW (black shaded contour). The HASSW shows a broader distribution compared to both low and intermediate $|\sigma_C|$ SSWs, overlapping with distributions observed for both FSW and regular SSW, implying that HASSW may have multiple source regions. Furthermore, these trends contradict those observed in Wind data, suggesting a possible radial evolution of Alfvénic SSW.

\subsection{Helium Abundance \label{sec:Windcompare}}
\begin{figure}
\epsscale{0.7}
\plotone{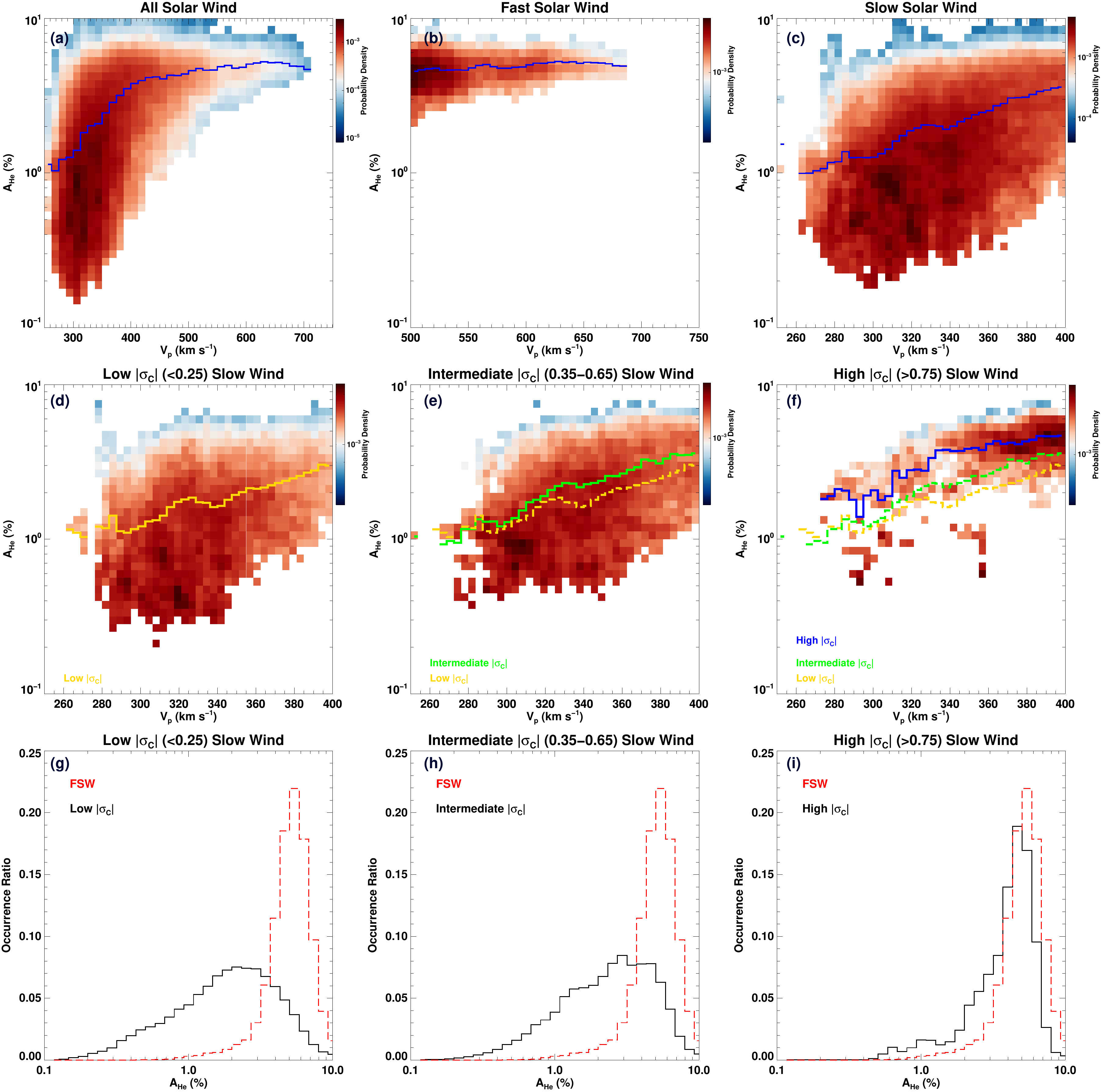}
\caption{Wind observations of helium abundance features in different solar wind streams. Panels (a) to (c) compare the helium abundance ($A_{He}$) variations with solar wind speed in all solar wind, fast solar wind, and slow solar wind, respectively. Panels (d) to (f) present $A_{He}$ in the slow solar wind with low, intermediate, and high Alfvénicities. The overplotted color line in each panel indicates the average $A_{He}$ for each speed bin. In panels (e-f), the average $A_{He}$ of slow wind with other Alfvénicities are overplotted for comparison, as indicated by the legend. The colors represent the probability density of $A_{He}$ within each bin. Panels (g) to (i) display the occurrence ratio of $A_{He}$ in the three types of slow solar wind, and the $A_{He}$ of FSW is overplotted with a red dashed line for comparison.} 
\label{fig:WHe_Wind}
\end{figure}

\begin{figure}
\epsscale{0.7}
\plotone{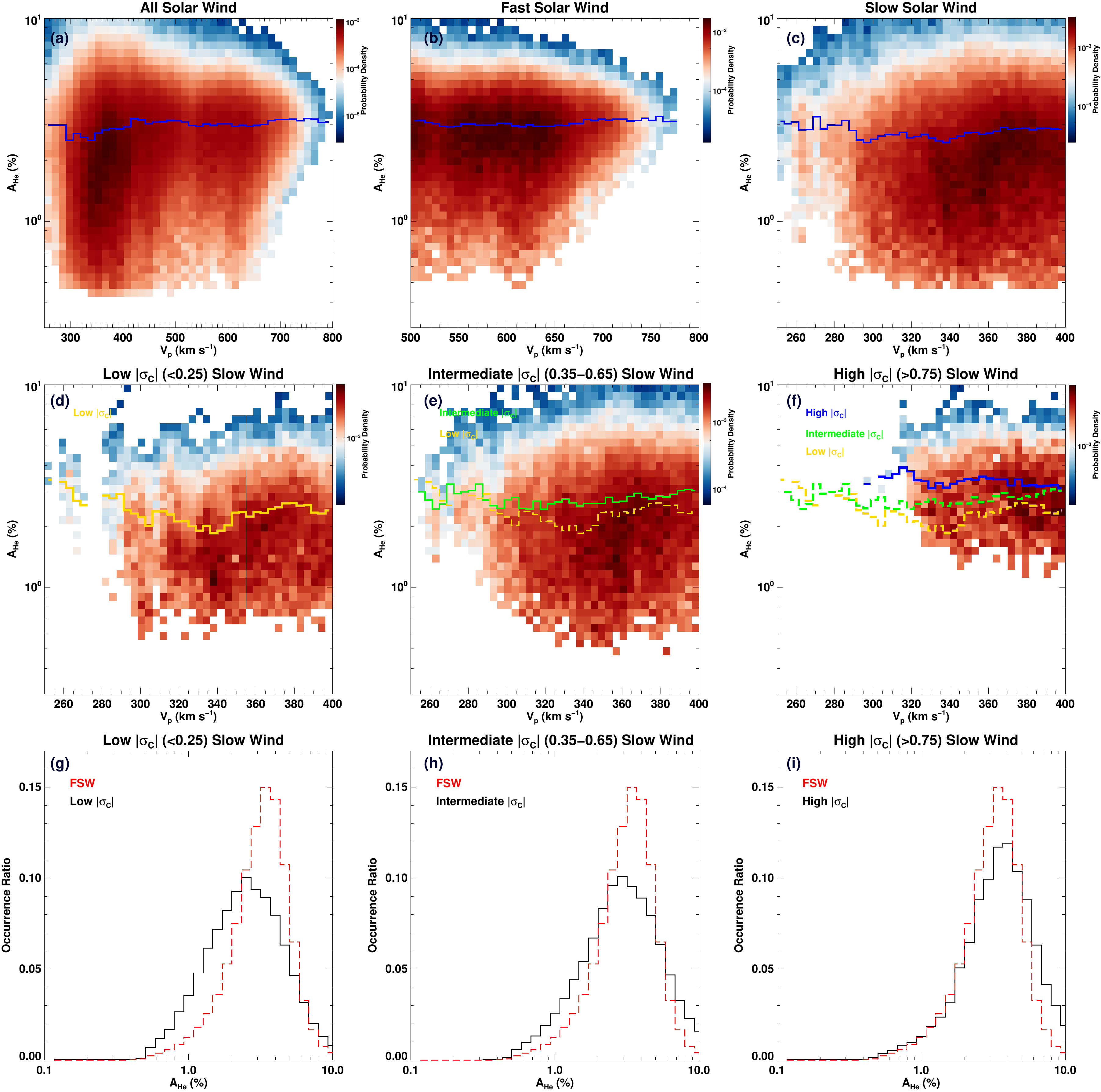}
\caption{Helios observations of helium abundance features in different solar wind streams. The format is the same as Figure \ref{fig:WHe_Wind}. } 
\label{fig:WHe_Helios}
\end{figure}

\begin{figure}
\epsscale{0.7}
\plotone{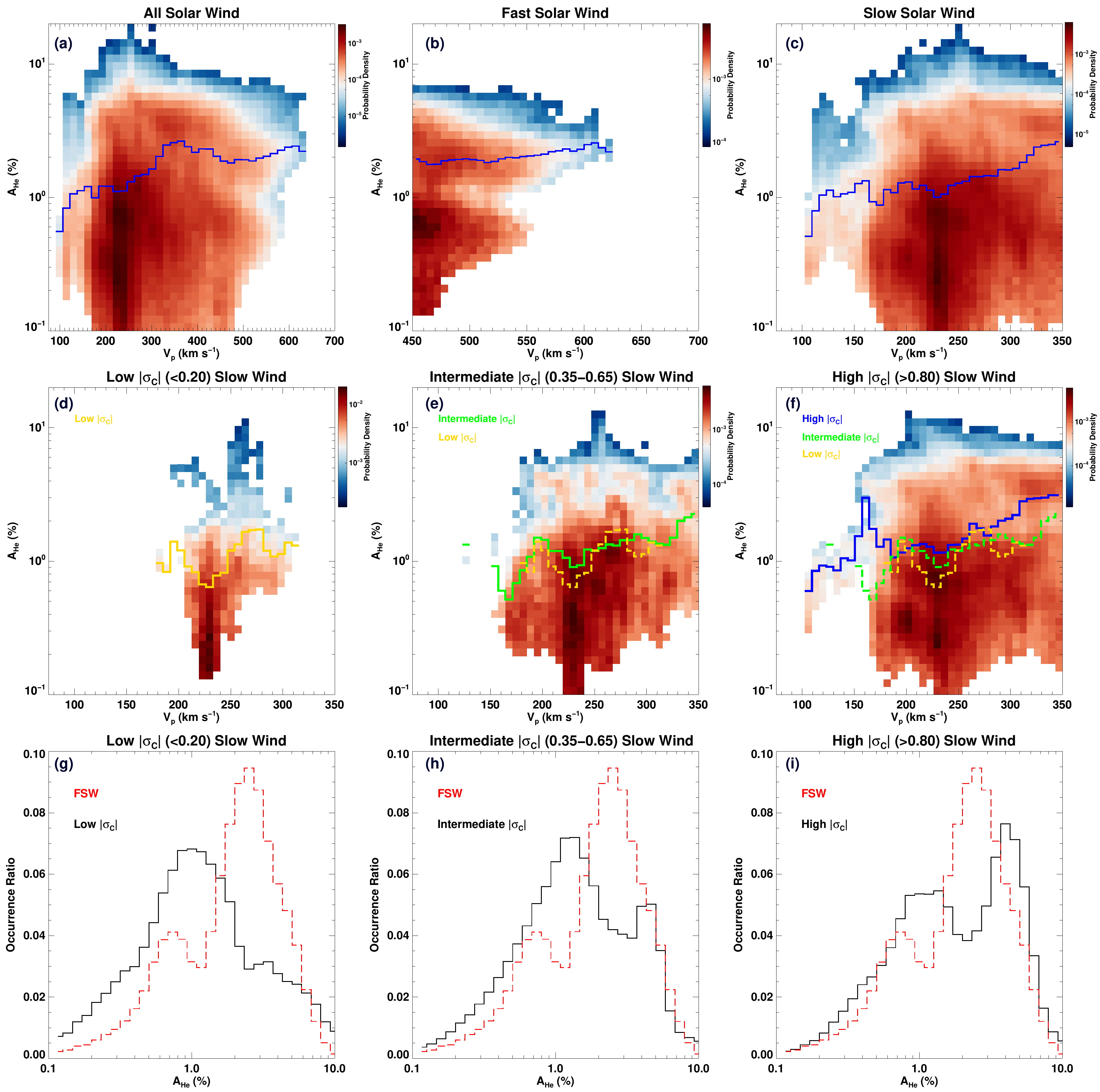}
\caption{PSP observations of helium abundance features in different solar wind streams. The format is the same as Figure \ref{fig:WHe_Wind}. } 
\label{fig:WHe_PSP}
\end{figure}

In this section, we further compare the helium abundance ($A_{He}$) features in different types of solar wind using Wind, Helios, and PSP observations.

Figure \ref{fig:WHe_Wind} illustrates the $A_{He}$ features in different solar wind streams using Wind observations. 
Panels (a-c) shows the $A_{He}$ variation as a function of solar wind speed in all solar wind, FSW, and SSW, respectively. The results indicate that FSW is predominantly characterized by helium-rich populations, with $A_{He}\approx4.5\%$. In contrast, the $A_{He}$ ratio in SSW shows an increasing trend with rising solar wind speed. Typically, FSW is dominated by helium-rich populations, while SSW consists of both helium-rich and helium-poor ($A_{He}\approx1.5\%$) populations, as reported by \citet{kasper-2007, kasper-2012} from long-term Wind observations. In SSW, the helium-poor population, which dominates during solar minimum, likely originates from streamer belt, whereas the helium-rich population, more prevalent during solar maximum, may come from active regions \citep{kasper-2007, kasper-2012} and/or small coronal holes \citep{wang-2017}. 
Panels (d-f) present the $A_{He}$ variations in SSW with low, intermediate, and high Alfvénicities. All three show an increasing trend with solar wind speed, with the average $A_{He}$ shifting to relatively higher values from low to high $|\sigma_C|$ SSW. Notably, panel (f) reveals two populations, with the helium-rich population dominating above 320 km/s and the helium-poor population below 320 km/s.
Panels (g-i) further compare the occurrence ratio of $A_{He}$ in the three types of SSW (black solid line) with that of FSW (red dashed line). The occurrence ratio of $A_{He}$ displays a wide spread covering both helium-rich and helium-poor populations, with the peak shifting from about 2\% in low $|\sigma_C|$ SSW to about 4.5\% in high $|\sigma_C|$ SSW. Given that HASSW likely undergoes different heating processes compared to FSW as indicated above, the helium-rich population of HASSW is probably sourced from active regions, though further investigation is needed to confirm this. However, the two populations in HASSW imply that the HASSW may have multiple source regions.

Figure \ref{fig:WHe_Helios} displays the $A_{He}$ variations using Helios data, following the same format as Figure \ref{fig:WHe_Wind}. 
Panels (a-c) show that FSW is dominated by helium-rich populations with $A_{He} \approx 3.5\%$, while SSW exhibits a similar but slightly lower $A_{He}$. However, these features are less robust in Helios observations compared to Wind data. Additionally, panels (d-f) indicate that the average $A_{He}$ shifts to relatively higher values in higher $|\sigma_C|$ SSW, although the overall variations are similar across SSW with different Alfvénicities. This pattern is also evident in the occurrence ratio of $A_{He}$ in the three types of SSW shown in panels (g-i). Consequently, Helios observations do not reveal a clear distinction between HASSW and FSW, which may be caused by the fact that the SSW recorded by Helios primarily originated from active regions and/or small coronal holes, or the Alfvénicity decreases with distance, as well as the possible mixing of SSW and FSW. 

Moreover, Figure \ref{fig:WHe_PSP} presents the $A_{He}$ signatures using PSP data, following the same format as Figure \ref{fig:WHe_Wind}. Similarly, panels (a-c) illustrate that FSW is dominated by helium-rich populations but with an overall lower $A_{He}$ being around 2.5\%, while SSW shows an increasing $A_{He}$ ratio with rising solar wind speed. Panels (d-f) exhibit similar variations across the different $|\sigma_C|$ SSWs, with the average $A_{He}$ shifting slightly to higher values in higher $|\sigma_C|$ SSW. However, panels (g-i) reveal robust features of the $A_{He}$ occurrence ratio in the three types of SSW. These panels clearly show that FSW consists of two populations: a major one peaking around $A_{He} \approx 2.5\%$ and a minor one peaking around $A_{He} \approx 0.7\%$. The helium-poor population within the FSW is likely associated with switchbacks, as they are generally characterized by a velocity spike \citep{kasper-2019} but a relatively low $A_{He}$ ratio as indicated by \citet{huang-2023SWBOrigin} based on the switchbacks identified in the first eight encounters of PSP mission. Furthermore, the occurrence ratio of $A_{He}$ in SSW distinctly manifests both a helium-poor population peaking around $A_{He} \approx 1.0\%$ and a helium-rich population peaking around $A_{He} \approx 4.0\%$, with the helium-rich population being more dominant in higher $|\sigma_C|$ SSW. It is very intriguing to observe that the helium-rich population in FSW appears to be situated between the helium-poor and helium-rich populations of SSW, and the mismatch between them further implies that they are from different source regions. This feature deserves a comprehensive study to uncover the helium particle characteristics in the pristine solar wind observed by PSP. Nevertheless, this result clearly indicates that the HASSW comprises two distinct populations.

\subsection{Radial Evolution} \label{sec:evo}

\begin{figure}
\epsscale{1.2}
\plotone{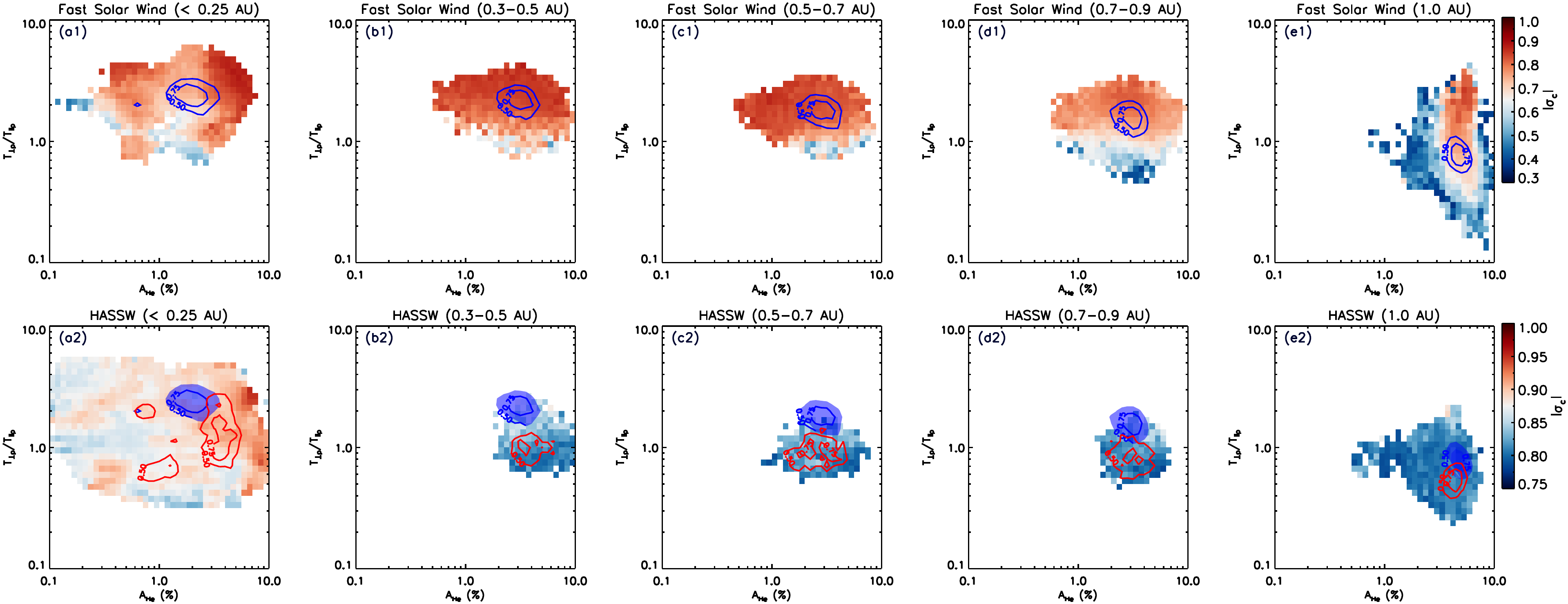}
\caption{Comparison of temperature anisotropy, helium abundance, and Alfvénicity between fast solar wind and high Alfvénicity slow solar wind at different heliocentric distances. Panels (a1) to (e1) show the distributions in the fast solar wind at distances below 0.25 AU, from 0.3 to 0.5 AU, from 0.5 to 0.7 AU, from 0.7 to 0.9 AU, and at around 1 AU, respectively, whereas panels (a2) to (e2) present the same distributions in high Alfvénicity slow solar wind. 
The color in each panel represents the median Alfvénicity in each bin. The blue contours in top panels indicate the 50\% and 75\% measurement levels of fast wind, which are correspondingly overplotted with blue shaded contours in bottom panels for comparison. The red contours in bottom panels indicate the 50\% and 75\% measurement levels of high Alfvénicity slow solar wind.}
\label{fig:evolution}
\end{figure}

The different features of temperature anisotropy and helium abundance at various heliocentric distances suggest a potential radial evolution of HASSW. In Figure \ref{fig:evolution}, we compare the temperature anisotropy, helium abundance, and Alfvénicity between FSW (top panels) and HASSW (bottom panels) at various distances from the Sun. The results below 0.25 AU, between 0.3 AU and 0.9 AU, and at around 1 AU are derived from PSP, Helios, and Wind observations, respectively. 

The top panels illustrate several characteristics of FSW. First, as heliocentric distance increases, the primary distribution transitions from anisotropic temperatures and relatively low $A_{He}$ to isotropic temperatures and higher $A_{He}$, as shown from panel (a1) to panel (e1). The isotropic temperatures are believed to be regulated by Coulomb collisions \citep{kasper-2008}, whereas the relatively low $A_{He}$ in PSP observations is likely related to low $A_{He}$ switchbacks \citep{huang-2023SWBOrigin}. Second, the anisotropic temperatures are often associated with relatively higher $|\sigma_C|$, confirming the FSW is characterized by an anisotropic Alfvénic feature. Third, the distribution of relatively higher $|\sigma_C|$ shifts from a broader spread over $A_{He}$ near the Sun to a narrower spread around $A_{He}\approx 4.5\%$ at 1 AU, implying a more significant disappearance of Alfvénicity in low $A_{He}$ fast streams. 
Besides, some of the fast wind at 1 AU shows very low $|\sigma_C|$ as well as low $A_{He}$, they may come from cold materials within interplanetary coronal mass ejections that are not well removed from this statistical study.

The bottom panels present the evolution of HASSW. It is evident that the overall Alfvénicity decreases significantly with increasing distance from the Sun, which may explain why PSP observed prevalent HASSW but Wind only detected limited occurrences. Additionally, HASSW below 0.25 AU exhibits a much broader spread in both $A_{He}$ and temperature anisotropy compared to HASSW at greater distances. For Helios and Wind observations, HASSW is predominantly composed of a high $A_{He}$ population, although a low $A_{He}$ population is present but constitutes a minor contribution. However, HASSW observed by PSP is distinctly characterized by two dominant populations. Moreover, the broad spread of temperature anisotropies within HASSW below 0.25 AU may be associated with both the origins of HASSW and the heating processes in the near-Sun environment. Furthermore, the comparison between HASSW (red contours) and FSW (blue dashed contours) reveals a reducing overlap from 1 AU to PSP's distance. This feature strongly suggests that HASSW and FSW may not share the same source regions.

\section{Discussion} \label{sec:dis}

\begin{figure}
\epsscale{1.}
\plotone{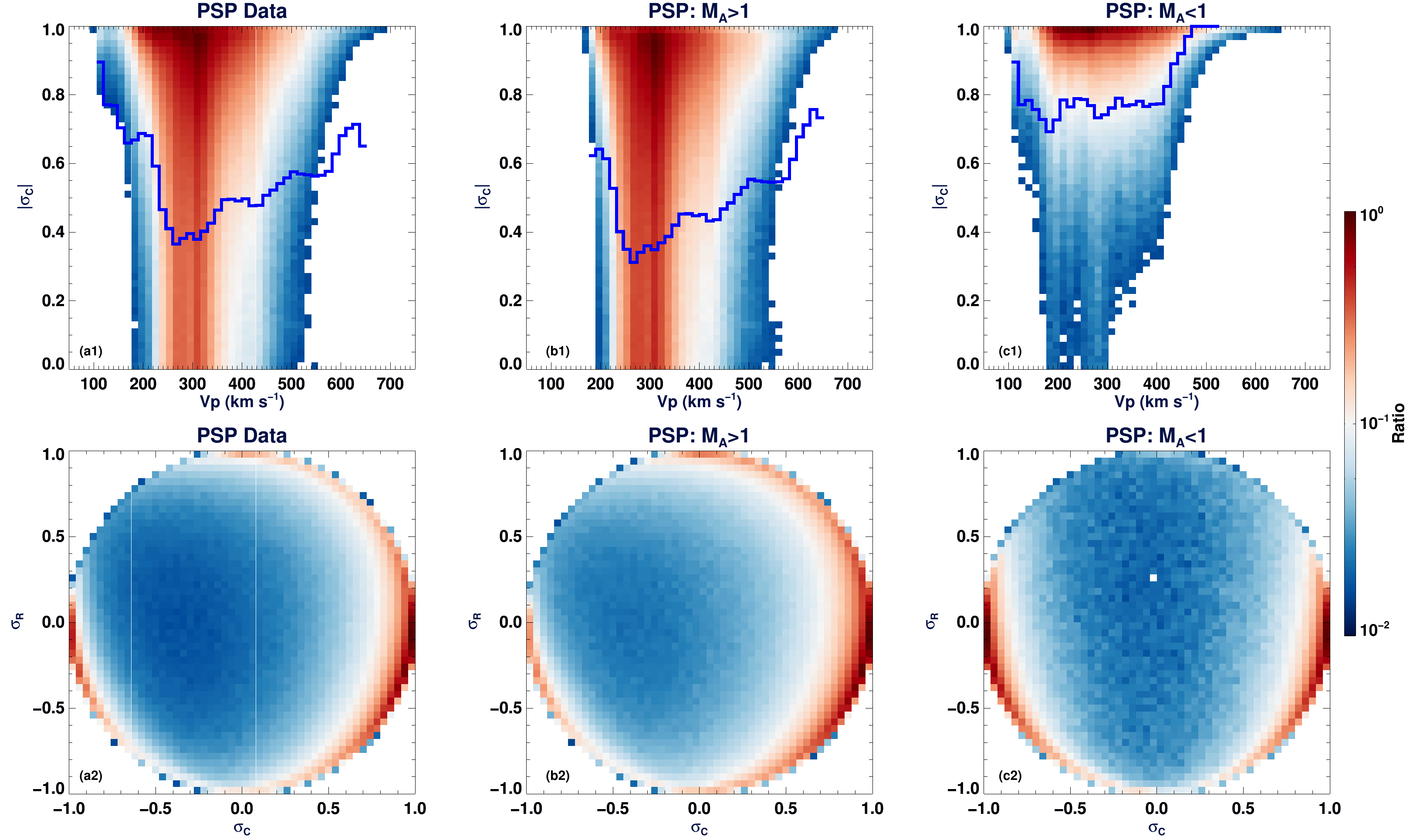}
\caption{PSP observations of absolute cross helicity variations in different solar wind streams. The format is the same as Figure \ref{fig:sgC}. From left to right, the panels present the variations in all solar wind, super-Alfvénic and sub-Alfvénic solar wind observed by PSP, respectively.   } 
\label{fig:sgc_comp}
\end{figure}

\begin{figure}
\epsscale{0.7}
\plotone{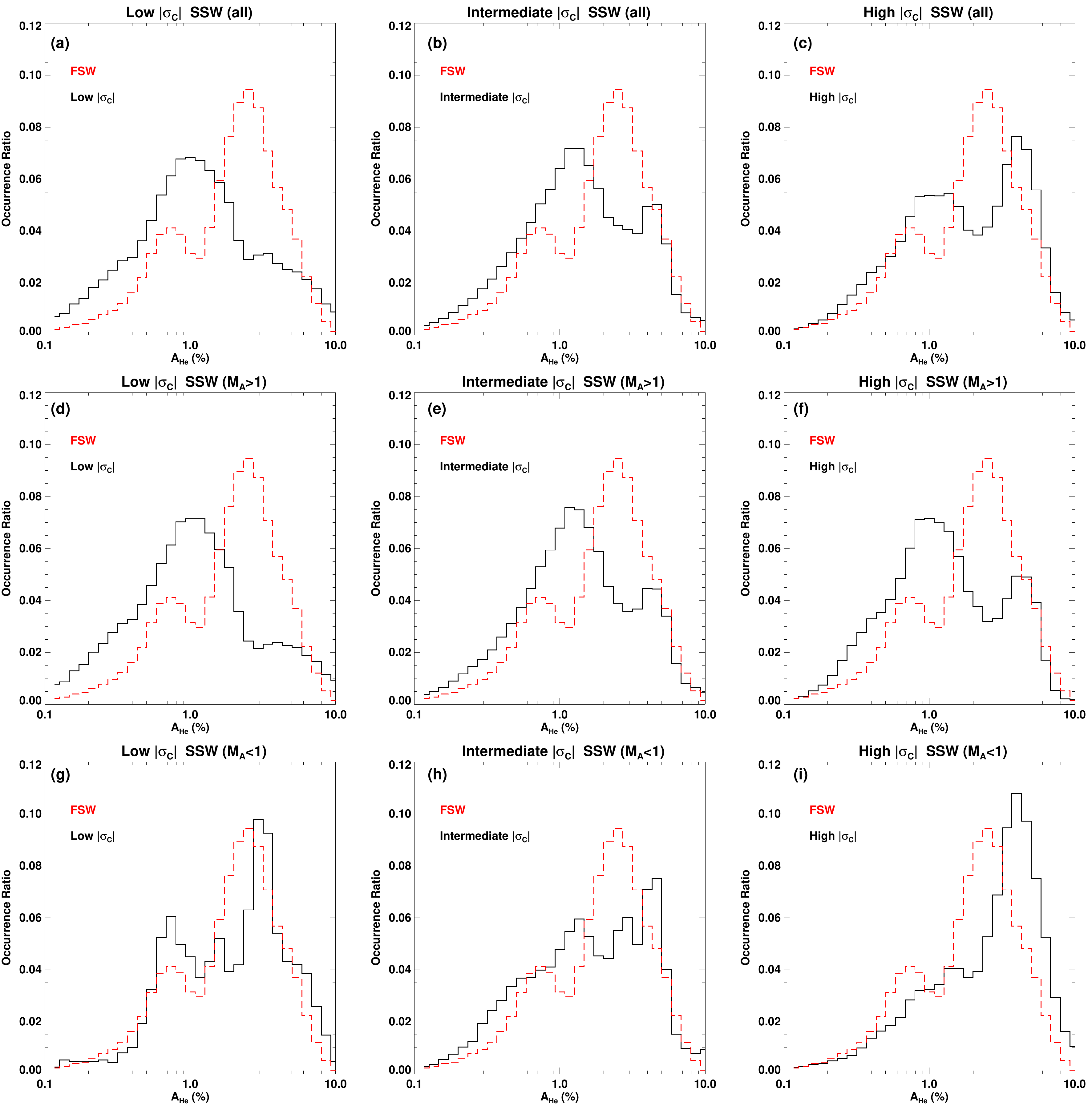}
\caption{ PSP observations of helium abundance occurrence ratio in different solar wind streams. Panels (a) to (c) compare the $A_{He}$ variations in all SSW with low, intermediate, and high Alfvénicities, respectively. Panels (d-f) and (g-i) present the $A_{He}$ distributions in super-Alfvénic and sub-Alfvénic SSW with low, intermediate, and high Alfvénicities, respectively. In each panel, the red dashed line represents the $A_{He}$ of FSW. } 
\label{fig:WHe_comp}
\end{figure}

\begin{figure}
\epsscale{0.7}
\plotone{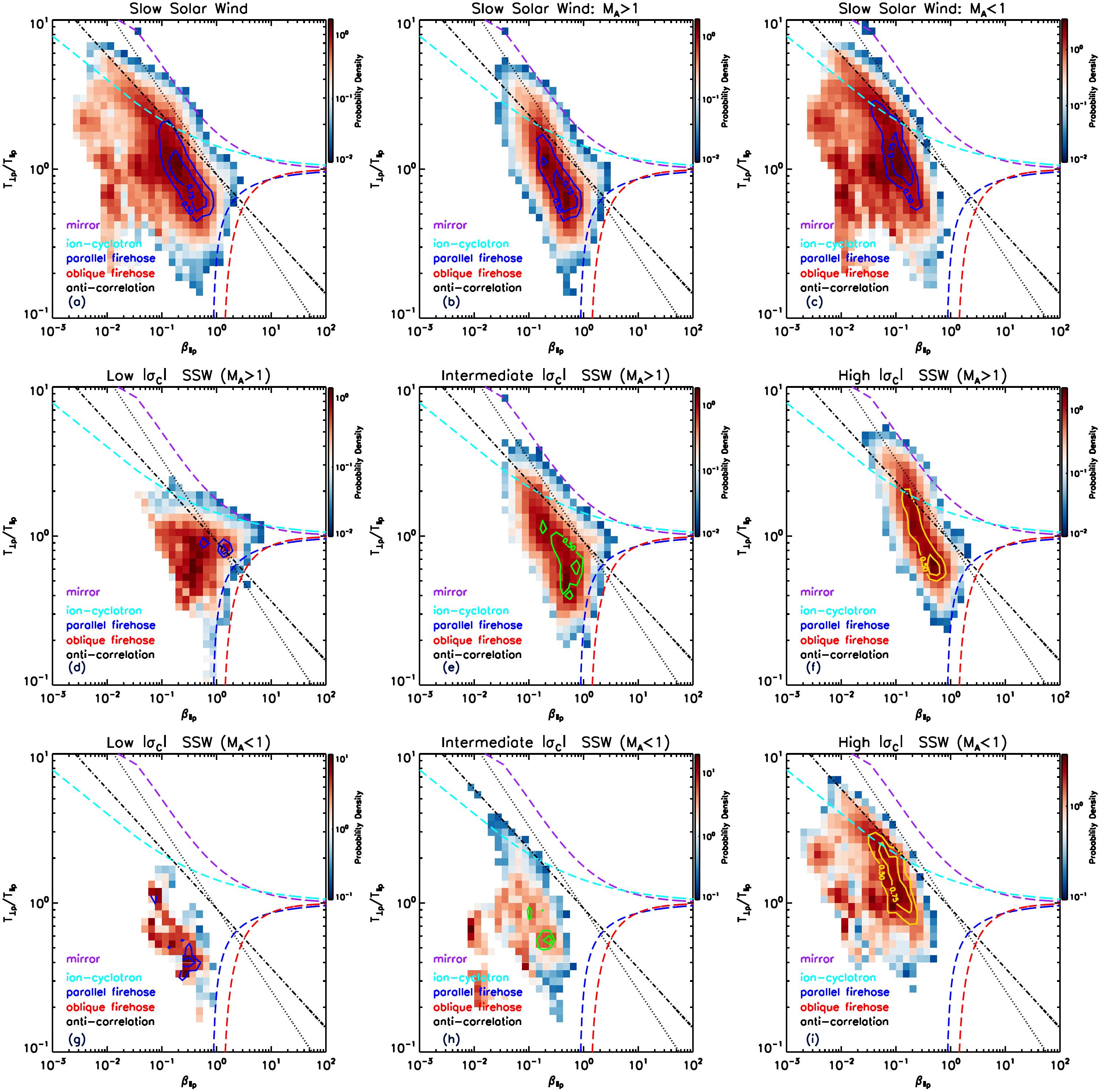}
\caption{ PSP observations of the probability density variations of temperature anisotropy versus parallel plasma beta in different solar wind streams. Panels (a) to (c) compare the temperature anisotropy variations in all SSW, super-Alfvénic and sub-Alfvénic SSW, respectively. Panels (d-f) and (g-i) present their distributions in super-Alfvénic and sub-Alfvénic SSW with low, intermediate, and high Alfvénicities, respectively. The lines in each panel are indicated by the legend and are same to those in Figure \ref{fig:TaniSSW_PSP}. } 
\label{fig:Tani_comp}
\end{figure}

As PSP crossed the Alfvénic critical surface for the first time on 28 April 2021 \citep{kasper-2021}, it observes more sub-Alfvénic solar wind, which is defined as the solar wind with an Alfvén Mach number below 1 ($M_A < 1$), in subsequent encounters \citep[e.g., ][]{bandyopadhyay-2022, chhiber-2022alfzone}. Studies suggest that many sub-Alfvénic solar wind recorded by PSP are characterized as HASSW \citep{zhao-2022turwaves, zhao-2022turbu, thepthong-2024alfvenicity}. Therefore, it is valuable to compare the Alfvénic slow solar wind in both sub-Alfvénic and super-Alfvénic ($M_A > 1$) regimes, and determine whether the unusually low speed HASSW is exclusively sub-Alfvénic. 

Figure \ref{fig:sgc_comp} presents the cross helicity variations in all solar wind (a1-a2), super-Alfvénic solar wind (b1-b2), and sub-Alfvénic solar wind (c1-c2) observed by PSP, respectively, with the same format as Figure \ref{fig:sgC}. As shown by the blue lines in panels (b1) and (c1), the ``U-shaped” pattern exists in both super-Alfvénic and sub-Alfvénic regimes, and they further suggest that the unusually low speed HASSW is not exclusively defined by $M_A < 1$. Moreover, the sub-Alfvénic solar wind seems to show a roughly stable portion of high Alfvénicity population between about 130 km/s and 400 km/s, whereas the super-Alfvénic solar wind exhibits a nearly monotonically decreasing to 260 km/s and then monotonically increasing trend. The variations imply that the Alfvénicity may have an internal relationship with solar wind acceleration in the inner heliosphere, which is recently discussed by \citet{alterman-2025}. 

Figure \ref{fig:WHe_comp} compares the PSP observations of $A_{He}$ distributions in FSW with those in all SSW (panels (a-c)), super-Alfvénic SSW (panels (d-f)), and sub-Alfvénic SSW (panels (g-i)) with low (left panels), intermediate (middle panels), and high (right panels) Alfvénicities, respectively. In comparison with the single-peak distribution in Figure \ref{fig:WHe_Wind} and Figure \ref{fig:WHe_Helios}, from this figure, we can see that the double-peak distribution exists in both super-Alfvénic and sub-Alfvénic SSW, and the two peaks do not align with those of FSW (red line). 
Moreover, the helium-rich population increases from low $|\sigma_C|$ SSW to high $|\sigma_C|$ SSW in both regimes, although the percentages differ. The overall $A_{He}$ distributions in low $|\sigma_C|$ and intermediate $|\sigma_C|$ SSW are predominantly determined by super-Alfvénic SSW, which overwhelmingly constitutes 81.1\% and 85.8\% of low and intermediate $|\sigma_C|$ SSW, respectively. However, in high $|\sigma_C|$ SSW, the helium-rich population is primarily contributed by sub-Alfvénic SSW, while the helium-poor population is majorly from super-Alfvénic SSW, as shown in panels (f) and (i).
Additionally, Figure \ref{fig:Tani_comp} shows the temperature anisotropy distributions in the two regimes, with a similar format as Figure \ref{fig:WHe_comp}. However, the figure does not exhibit notable differences between super-Alfvénic SSW (panels (d-f)) and sub-Alfvénic SSW (panels (g-i)), except that the parallel plasma beta is relatively smaller in sub-Alfvénic SSW. 
Consequently, these results suggest that the two populations in HASSW are not independently contributed by sub-Alfvénic SSW and super-Alfvénic SSW, and the HASSW should originate from multiple source regions. 

From panels (f) and (i) in Figure \ref{fig:WHe_comp}, we can see that super-Alfvénic HASSW constitutes more helium-poor population ($A_{He}\approx 1\%$), whereas sub-Alfvénic HASSW contains more helium-rich population ($A_{He}\approx 4\%$). This result implies that sub-Alfvénic HASSW may predominantly originate from small coronal holes, coronal hole boundaries, or active regions \citep[e.g., ][]{kasper-2007, stansby-2019, bale-2019, ervin-2024comp, ervin-2024sasw}, while super-Alfvénic HASSW may primarily come from closed field regions. The shift in the two helium populations from sub-Alfvénic to super-Alfvénic HASSW infers that Alfvénicity may decrease more rapidly in sub-Alfvénic HASSW with a helium-rich population, likely due to strong solar wind interactions, mixing, and large velocity shears therein \citep{zhou-1989WKB, bruno-2006, dAmicis-2018, thepthong-2024alfvenicity}.     

In addition, these findings raise several intriguing questions for further investigation. 
First, what is primary factor driving the degradation of Alfvénicity in different types of solar wind? The prevalence of HASSW in the inner heliosphere implies that most solar wind likely originates with high Alfvénicity, but the Alfvénicity decreases rapidly in slow solar wind or farther from the Sun. The primary factors are yet to be determined. For example, the mixing of inward and outward propagating Alfvénic fluctuations can significantly reduce Alfvénicity, particularly around heliospheric current sheets (HCSs), while the interactions between different solar wind streams can greatly diminish Alfvénicity in their transition regions. So it is valuable to identify the key factor that governs Alfvénicity variations in, such as, very low speed HASSW and sub-Alfvénic solar wind, and determine the possible relationship between Alfvénicity variation and solar wind acceleration.    
Second, what are the underlying differences in heating mechanisms between FSW and HASSW?  Their differing temperature anisotropies suggest distinct heating processes. In general, temperature anisotropies can be regulated by Coulomb collisions, instabilities, wave activities, turbulence, and other factors \citep[e.g., ][]{kasper-2008, bale-2009, verscharen-2016}. Determining the underlying heating mechanisms will provide valuable insights into the processes driving heating in different solar wind environments. 
Third, why are the $A_{He}$ distributions anti-correlated in HASSW and FSW? As shown in Figure \ref{fig:WHe_comp}, the double-peak $A_{He}$ distributions appear anti-correlated in FSW (peaks at 0.7\% and 2.5\%) and HASSW (peaks at 1\% and 4\%). It is unclear why FSW exhibits even lower $A_{He}$ values than these SSW, which requires further exploration. 

\section{Summary} \label{sec:sum}

In this work, we conduct a statistical analysis of the temperature anisotropy and helium abundance characteristics of HASSW using data from PSP below 0.25 AU, Helios between 0.3 AU and about 1 AU, and Wind around 1 AU. Our findings reveal several outstanding features of HASSW:

\begin{enumerate}
\item \textbf{Prevalent existence of HASSW in the inner heliosphere.} The Alfvénicity as a function of solar wind speed shows a monotonic increasing trend in both Helios and Wind observations. However, in PSP observations, it exhibits a ``U-shaped” pattern, highlighting a distinct population of unusually low speed HASSW, and the ``U-shaped” pattern exists in both sub-Alfvénic and super-Alfvénic regimes. Moreover, high Alfvénicity populations account for 44.1\%, 46.4\%, and 33.6\% of all solar wind measured by PSP, Helios, and Wind, respectively. Among there, HASSW dominates 81.8\%, 35.1\%, and 58.6\% of the high Alfvénicity populations for PSP, Helios, and Wind, respectively, underscoring the prevalent existence of HASSW in the inner heliosphere.

\item \textbf{Evolving overlaps of temperature anisotropy distributions between HASSW and FSW.} The overlap in temperature anisotropy distributions between HASSW and FSW decreases with increasing heliocentric distance, suggesting that HASSW may undergo different heating processes compared to FSW. 

\item \textbf{Two populations of helium abundance in HASSW.} HASSW recorded by Helios and Wind shows evidence of two $A_{He}$ populations, with the overall average $A_{He}$ being higher than that in SSW with low and intermediate Alfvénicities and being closer to the values observed in FSW. 
However, PSP measurements clearly indicate that HASSW comprises two distinct populations: a helium-rich population primarily from sub-Alfvénic HASSW, and a helium-poor population mainly from super-Alfvénic HASSW. Moreover, it is noteworthy that the two $A_{He}$ peaks are anti-correlated with those of FSW. 

\item \textbf{Radial evolution highlights differences between HASSW and FSW.} The comparison of HASSW and FSW from 1 AU to PSP’s closest approach to date reveals a decreasing overlap in their temperature anisotropy versus $A_{He}$ space. This robust feature indicates that some of the HASSW likely originates from different source regions from FSW, and they may evolve differently as they propagate outward. In FSW, the $A_{He}$ distribution narrows from a broader spread near the Sun to a concentrated value around $4.5\%$ at 1 AU, implying a more significant disappearance of Alfvénicity in helium-poor fast streams. However, the shift from a helium-rich to a helium-poor dominant population as HASSW transits from sub-Alfvénic to super-Alfvénic regime infers that Alfvénicity may decrease more rapidly in helium-rich, sub-Alfvénic HASSW. 
\end{enumerate}

In conclusion, our statistical results highlight non-ignorable differences between HASSW and FSW, particularly in temperature anisotropy and helium abundance variations, strongly suggesting that not all HASSW originate from the same source regions as FSW.

\acknowledgments
We thank the reviewer for their constructive suggestions, which have helped improve this work.
This work is supported by NASA grant 80NSSC23K0737. Parker Solar Probe was designed, built, and is now operated by the Johns Hopkins Applied Physics Laboratory as part of NASA’s Living with a Star (LWS) program (contract NNN06AA01C). Support from the LWS management and technical team has played a critical role in the success of the Parker Solar Probe mission. We acknowledge the use of Helios data from the Helios data archive (\url{http://helios-data.ssl.berkeley.edu/}), the Parker Solar Probe and Wind data from SPDF/CDAWeb (\url{https://spdf.gsfc.nasa.gov/pub/data/}), and the sunspot data from the World Data Center SILSO, Royal Observatory of Belgium, Brussels.

\bibliography{SSW}{}
\bibliographystyle{aasjournal}

\end{document}